\documentclass[twocolumn,tighten]{aastex61}

\submitjournal{AAS Journals}
\received{14 March 2017}
\revised{3 April 2017}

\newcommand{\obj}{2013 SY$_{99}$}
\newcommand{\mina}{694} 
\newcommand{\maxa}{778} 
\newcommand{\besta}{733} 
\newcommand{\sigmaa}{42} 

\shorttitle{OSSOS V: a high-perihelion distant Solar System object}
\shortauthors{Bannister et al.}

\begin{document}

\title{OSSOS: V. Diffusion in the orbit of a high-perihelion distant Solar System object} 

\author[0000-0003-3257-4490]{Michele T. Bannister}
\correspondingauthor{Michele T. Bannister}
\email{michele.t.bannister@gmail.com}
\affiliation{Astrophysics Research Centre, Queen's University Belfast, Belfast BT7 1NN, United Kingdom}
\affiliation{NRC-Herzberg Astronomy and Astrophysics, National Research Council of Canada, 5071 West Saanich Rd, Victoria, British Columbia V9E 2E7, Canada}
\affiliation{Department of Physics and Astronomy, University of Victoria, Elliott Building, 3800 Finnerty Rd, Victoria, BC V8P 5C2, Canada}

\author[0000-0002-3507-5964]{Cory Shankman}
\affiliation{Department of Physics and Astronomy, University of Victoria, Elliott Building, 3800 Finnerty Rd, Victoria, BC V8P 5C2, Canada}

\author[0000-0001-8736-236X]{Kathryn Volk}
\affiliation{Lunar and Planetary Laboratory, University of Arizona, 1629 E University Blvd, Tucson, AZ 85721, United States}

\author[0000-0001-7244-6069]{Ying-Tung Chen}
\affiliation{Institute of Astronomy and Astrophysics, Academia Sinica; 11F of AS/NTU Astronomy-Mathematics Building, Nr. 1 Roosevelt Rd., Sec. 4, Taipei 10617, Taiwan, R.O.C.}

\author{Nathan Kaib}
\affiliation{HL Dodge Department of Physics \& Astronomy, University of Oklahoma, Norman, OK 73019, United States}

\author{Brett J. Gladman}
\affiliation{Department of Physics and Astronomy, University of British Columbia, Vancouver, BC, Canada}

\author[0000-0002-4385-1169]{Marian Jakubik}
\affiliation{Astronomical Institute, Slovak Academy of Science, 05960 Tatranska Lomnica, Slovakia}

\author[0000-0001-7032-5255]{J. J. Kavelaars}
\affiliation{NRC-Herzberg Astronomy and Astrophysics, National Research Council of Canada, 5071 West Saanich Rd, Victoria, British Columbia V9E 2E7, Canada}
\affiliation{Department of Physics and Astronomy, University of Victoria, Elliott Building, 3800 Finnerty Rd, Victoria, BC V8P 5C2, Canada}

\author{Wesley C. Fraser}
\affiliation{Astrophysics Research Centre, Queen's University Belfast, Belfast BT7 1NN, United Kingdom}

\author[0000-0003-4365-1455]{Megan E. Schwamb}
\affiliation{Gemini Observatory, Northern Operations Center, 670 North A'ohoku Place, Hilo, HI 96720, United States}

\author{Jean-Marc Petit}
\affiliation{Institut UTINAM UMR6213, CNRS, Univ. Bourgogne Franche-Comt\'e, OSU Theta F25000 Besan\c{c}on, France}

\author{Shiang-Yu Wang} 
\affiliation{Institute of Astronomy and Astrophysics, Academia Sinica; 11F of AS/NTU Astronomy-Mathematics Building, Nr. 1 Roosevelt Rd., Sec. 4, Taipei 10617, Taiwan, R.O.C.}

\author{Stephen D. J. Gwyn}
\affiliation{NRC-Herzberg Astronomy and Astrophysics, National Research Council of Canada, 5071 West Saanich Rd, Victoria, British Columbia V9E 2E7, Canada}

\author[0000-0003-4143-8589]{Mike Alexandersen} 
\affiliation{Institute of Astronomy and Astrophysics, Academia Sinica; 11F of AS/NTU Astronomy-Mathematics Building, Nr. 1 Roosevelt Rd., Sec. 4, Taipei 10617, Taiwan, R.O.C.}

\author{Rosemary E. Pike}
\affiliation{Institute of Astronomy and Astrophysics, Academia Sinica; 11F of AS/NTU Astronomy-Mathematics Building, Nr. 1 Roosevelt Rd., Sec. 4, Taipei 10617, Taiwan, R.O.C.}

\begin{abstract}

We report the discovery of the minor planet \obj, on an exceptionally distant, highly eccentric orbit. With a perihelion of 50.0~au, \obj's orbit has a semi-major axis of $730 \pm 40$~au, 
the largest known for a high-perihelion trans-Neptunian object (TNO), well beyond those of (90377) Sedna and 2012 VP$_{113}$. Yet, with an aphelion of $1420 \pm 90$~au, \obj's orbit is interior to the region influenced by Galactic tides.
Such TNOs are not thought to be produced in the current known planetary architecture of the Solar System, and they have informed the recent debate on the existence of a distant giant planet.
Photometry from the Canada-France-Hawaii Telescope, Gemini North and Subaru indicate \obj\ is $\sim 250$~km in diameter and moderately red in colour, similar to other dynamically excited TNOs.
Our dynamical simulations show that Neptune's weak influence during \obj's perihelia encounters drives diffusion in its semi-major axis of hundreds of astronomical units over 4~Gyr.
The overall symmetry of random walks in semi-major axis allow diffusion to populate \obj's orbital parameter space from the 1000--2000~au inner fringe of the Oort cloud. 
Diffusion affects other known TNOs on orbits with perihelia of 45 to 49~au and semi-major axes beyond 250~au, 
providing a formation mechanism that implies an extended population, gently cycling into and returning from the inner fringe of the Oort cloud.

\end{abstract}


\section{Introduction}
\label{sec:intro}

The distant Solar System contains a number of small bodies on orbits that are difficult to produce with the currently known planetary architecture.
Certain trans-Neptunian objects (TNOs) have orbits with semi-major axes $a \gtrsim 50$ astronomical units (au) that are ``detached'' from present gravitational interaction with Neptune. 
These detached objects have perihelia $q \gtrsim 37$ au \citep{Lykawka:2007ff,Gladman2008}, 
much more distant than Neptune's $\sim 30$ au orbit, which prevents substantial gravitational interaction with Neptune even when they are closest to the Sun; they are also not in resonance with Neptune.

The zone of detached orbits with moderate $a \sim 50-250$ au is finely filigreed with high-order mean-motion resonances with Neptune, and simulations show minor planets could have been emplaced with low efficiency in the era of Neptune's migration \citep{Lykawka:2007Icar}, out as far as a soft boundary in the region $a \lesssim 250$ au.
Methods include resonance sticking in the scattering population or capture into a mean-motion resonance \citep{Brasser:2013dw, Pike2016nice}, followed by evolution on a periodic orbit of the third kind\footnote{Often termed ``MMR + Kozai", which is not the effect at work here: see discussion in \citet{Malhotra:2016ab}.} \citep{2005CeMDA..91..109G,Gallardo:2012Icar}.

For even larger-$a$ detached orbits, the emplacement history remains more mysterious. 
The distant semi-major axes of these TNOs are still much smaller than is typical for Oort cloud objects, whose dynamics are significantly affected by Galactic tides. 
The ``inner fringe" of the Oort cloud we consider as $a \sim 1000-2000$ au; orbits in this quiet zone have placid dynamical interactions, with tides, passing stars and planetary perturbations only becoming relevant on Gyr timescales. 
\citet{Brasser:2014iw} found TNOs with perihelia $q > 40$ AU and semi-major axes $250 < a \lesssim 2000$ au are currently isolated from gravitational interaction with Neptune or with Galactic tides. 
Similarly, \citet{Kaib:2009Sci} examined the production of long period comets and found that the production efficiency drops significantly for bodies with $a <3000$ au compared to those with larger semi-major axes. 
There is no clear consensus on where the Oort cloud dominates; we refer to it as $a \gtrsim 2000$ au \citep{Dones:2004,Kaib:2009Sci}.

Minor planets with high-perihelia orbits and $a \lesssim 2000$ au are thus only weakly affected by tides and stellar impulses in the current solar environment.
Only a few are known: merely six published TNOs have $q > 40$ and $a > 250$ au (Table~\ref{tab:orbits}).
We choose to refer to $q > 40$, $a > 250$ au minor planets as ``extreme" TNOs,
avoiding association with particular objects (which may or may not be fully representative of the class) and with formation mechanisms; the exact boundaries of the class are still being defined. 

The extreme dynamical class of TNOs are a signature of a yet-to-be-determined aspect of the Solar System's architecture or history \citep{Gladman:2002Icar,Brown:2004fp}. 
Candidate explanations include
stellar perturbations on the forming TNO population while the Sun was still in its birth cluster \citep{Brasser:2012gi},
capture of objects from another star in the birth cluster \citep{Kenyon:2004aa,Jilkova:2015gf},
changing proximity to other stars and changing tides as the Sun migrates within the Galaxy \citep{Kaib:2011ky},
gravitational sculpting of distant TNOs during a stellar flyby \citep{2004AJ....128.2564M,Kenyon:2004aa},
and perihelion lifting by an unseen exterior planet, 
either a ``rogue" during its departure from the system \citep{Gladman:2005Sci...307...71G,2006ApJ...643L.135G} 
or one still orbiting at present \citep{Gladman:2002Icar,Brown:2004fp,Gomes:2006en,Soares:2013hf,Trujillo:2014ih,Batygin:2016ef}.

The extreme TNO population spend just a small fraction of their orbits near their perihelia, within the magnitude limits attainable by large-aperture telescopes.
Analysis of the population is therefore particularly sensitive to discovery biases.
The proposed formation scenarios cannot yet be observationally distinguished.
\citet{Schwamb:2010p932,Brasser:2012gi,Trujillo:2014ih} find that the stellar cluster origin is consistent with the observed TNOs, but the other scenarios have not been ruled out.

In spite of the low observational probabilities, we recently discovered an exceptional new TNO. 
In the following sections, we show that \obj\ has an orbit with $q = 50.0 \pm 0.05$ au that has the largest semi-major axis yet found among extreme TNOs:
our four years of observation constrain the semi-major axis to be in the range $\mina < a < \maxa$ au.
Our observations also determine \obj's absolute magnitude ($H_r = 6.81\pm0.14$) and mildly red colour.
We model the dynamics of \obj\ in the known Solar System, and find that \obj\ is stable on Gyr timescales; however, its orbit is so weakly bound that very distant interactions with Neptune at perihelion drive a significant random walk in semi-major axis on timescales of tens of Myr, as in \citet{1987AJ.....94.1330D}.
We model the evolution of other extreme TNO orbits, and find that gentle evolution via diffusion is a dynamical pathway, linking the populations of the inner fringe of the Oort cloud and the extreme trans-Neptunian objects. 
We also model \obj's evolution in a Solar System containing a possible distant planet: \obj\ shows substantial orbital instability, or becomes unobservable.

\floattable
\begin{deluxetable*}{lLLLRRRRrLl}
\tabletypesize{\scriptsize}
\tablecaption{Barycentric orbital elements of trans-Neptunian objects with $q > 40$ and $a > 250$ au in the International Celestial Reference System at epoch MJD 57793\label{tab:orbits}}
\tablehead{
\colhead{Name} & \colhead{$q$} & \colhead{$a$} & \colhead{$e$} & \colhead{$i$} & \colhead{$\Omega$} & \colhead{$\omega$} & \colhead{$\varpi$} & \colhead{Arc} & \colhead{$H_r$} & \colhead{Discovery} \\ 
\colhead{} & \colhead{(au)} & \colhead{(au)} & \colhead{} & \colhead{(\degr)} & \colhead{(\degr)} & \colhead{(\degr)} & \colhead{(\degr)} & \colhead{(days)}  & \colhead{} & \colhead{} 
}
\startdata
2012 VP$_{113}$ & 80.3^{+1.2}_{-1.6}  		    & 266^{+26}_{-17}    	& 0.69   \pm 0.03 	& 24.1 & 90.8 & 292.7 	& 24.4 	& 739  & 4.0 & \citet{Trujillo:2014ih}  \\ 
(Sedna) 2003 VB$_{12}$ & 76.19 \pm 0.03 	    & 507 \pm 10 & 0.8496 \pm 0.003 & 11.9 & 144.4 & 311.3 	& 96.1 	& 9240 & 1.5 & \citet{Brown:2004fp}  \\ 
{\bf 2013 SY$_{99}$}	    & {\bf 50.0 \pm 0.1}	& {\bf \besta \pm \sigmaa} & {\bf 0.932 \pm 0.007 } & {\bf4.2} & {\bf 29.5 } & {\bf 32.2 }	&  {\bf  61.7  }& {\bf 1156 }& {\bf 6.81 \pm 0.14} & {\bf  this work} \\ 
2010 GB$_{174}$ & 48.6 \pm 0.1     		& 351 \pm 9 	& 0.862 \pm 0.004 	& 21.6 & 130.7 & 347.2 	& 118.5    & 965 & 6.5 & \citet{Chen:2013gs} \\ 
2014 SR$_{349}$ & 47.5 \pm 0.2 		    & 299 \pm 12 	& 0.841 \pm 0.007 	& 18.0 & 34.9 & 341.2 	& 16.3 	& 738 & 6.6 & \citet{Sheppard:2016jf}  \\ 
(474640) 2004 VN$_{112}$ & 47.321 \pm 0.004 	    & 316 \pm 1 	& 0.8505 \pm 0.0005 & 25.6 & 66.0 & 327.1 	& 33.1 	& 5821 & 6.5 & \citet{2008ApJ...682L..53B}  \\ 
2013 FT$_{28}$    & 43.47 \pm 0.08 		& 295 \pm 7 	& 0.853 \pm 0.004 	& 17.4 & 217.7 & 40.7 	& 258.0 	& 1089 &  6.7 & \citet{Sheppard:2016jf}  \\ 
\enddata
\tablecomments{
Uncertainties are from the Monte Carlo approach of \citet{Gladman2008}, and are shown in full for $q$, $a$, $e$. The much more precisely quantified angular elements are shown only to $0.1^\circ$.
}
\end{deluxetable*}

\section{Observations}
\label{sec:discovery}

\obj\footnote{Internal survey designation is \texttt{uo3l91}; the TNO has also been referred to as ``L91".} was found in observations at opposition on 2013 September 29 with the 0.9 deg$^{2}$ field of view MegaPrime imager \citep{2003SPIE.4841...72B} of the Canada-France-Hawaii Telescope (CFHT) on Maunakea.
Three images in the r.3901 filter were taken spanning two hours, targeting the 21 deg$^{2}$ ``13BL" survey area of the Outer Solar System Origins Survey (OSSOS).
This grid of 3 x 7 MegaPrime pointings centered on R.A. 0\fh54\arcmin, decl. +3\arcdeg50\arcmin\ was the third of the eight target regions of OSSOS. 
The observation and data analysis techniques for the survey are detailed in \citet{Bannister:2016cp}.

OSSOS completely recovers all TNOs discovered in each target region that are brighter than a limiting magnitude --- that region's \textit{characterization limit}. 
89 TNOs were discovered in 13BL; 10 were fainter than the characterization limit, including \obj, which had a mean $m_r = 24.8 \pm 0.3$ at discovery.
For \obj's rate of apparent sky motion (2.14 \arcsec/hr), the characterization limit of the discovery observations was $m_r = 24.45$. Extrapolating sky motion and magnitude into the less understood \textit{uncharacterized} region of indeterminate detection efficiency, \obj\ was found with a detection efficiency of $\sim 0.09$ \citep[Eqn. 2]{Bannister:2016cp}.
Faint, uncharacterized objects are very difficult to recover and track: they are only imaged if they serendipitously fall within the wide MegaPrime field of view during observations to refine the orbits of the brighter characterized TNOs.
The four months of serendipitous observations of \obj\ in 2013 implied that it was plausibly on an actively scattering orbit. 
However, \obj's barycentric discovery distance of 60 au, substantially greater than the $25 < d < 52$ au discovery distance of most OSSOS TNOs, flagged it as potentially unusual.
We thus purposefully recovered \obj\ two years later.
The orbital arc for \obj\ is presently described by a dense and well-sampled set of 33 observations from 2013 September 5 to 2016 November 4 (Table~\ref{tab:photometry}).
The astrometry\footnote{Future link to the Minor Planet Center MPEC.} is measured against the precisely calibrated OSSOS plate solution \citep{Bannister:2016cp}, with median residuals of only 0.1\arcsec.
Unfortunately, \obj\ is too faint to be visible in any publicly available archival imaging \citep{Gwyn:2012gv}.  

We measured the optical colours of \obj\ with imaging with the 8.1 m Frederick C. Gillett Gemini North Telescope and the 8.2 m Subaru Telescope, at high airmass (Table~\ref{tab:colours}).  
A non-consecutive $r$ and $i$ sequence was observed over four hours with Subaru's Hyper Suprime Cam (HSC) \citep{2012SPIE.8446E..0ZM} on 2016 January 10 (Table~\ref{tab:photometry}).
A consecutive $g$ and $r$ sequence\footnote{Fast Turnaround program GN-2015B-FT-26.} was observed over two and a half hours on 2016 January 11 with the Gemini Multi-Object Spectrograph (GMOS-N) imager \citep{2004PASP..116..425H} with 2x2 binning (Table~\ref{tab:photometry}).
All observations were with sidereal tracking, allowing photometric calibration to Sloan Digital Sky Survey (SDSS) \citep{SDSS_DR12:2015ApJS} stars.
Bias and flat field correction for the GMOS-N data was 
with the Gemini Observatory Ureka IRAF package, and for the HSC data with the package {\tt hscPipe}.
Due to the difficulty of merging the astrometric plate solutions to that of OSSOS from CFHT, the Subaru and Gemini observations do not contribute to our determination of \obj's orbit (\S\ref{sec:shortterm}).
Aperture photometry was measured with the moving-object photometry package TRIPPy \citep{Fraser:2016bd}, which includes point-spread function (PSF) fitting and subtraction.
We find no evidence for binarity of \obj.

The few measured optical reflectance colours of extreme TNOs range from mildly to considerably redder than solar\footnote{\url{http://sdss.org/dr12/algorithms/ugrizvegasun/}} $g-r = 0.44 \pm 0.02$ (Table~\ref{tab:colours}).
We corrected the Gemini instrumental magnitudes to the SDSS by colour terms derived in the {\it Colours of OSSOS} survey (Schwamb et al., in prep):
\begin{equation}
g_{G}=g_{S}-0.146 (\pm0.002) \left(g_{S}-r_{S}\right),
\end{equation}
\begin{equation}
r_{G}=r_{S}-0.048 (\pm0.003) \left(g_{S}-r_{S}\right).
\end{equation}
We similarly set the Subaru zeropoints to the SDSS, 
adopting an $i$-band colour term (Pike et al., in prep); while approximate for HSC, the uncertainty it introduces is significantly smaller than the Poisson uncertainty on the photometry.
\obj's $r-i=0.46 \pm 0.10$ is slightly red in $r-i$ for a small dynamically excited TNO, but still consistent at $1.5\sigma$ with the $r-i\sim 0.3$ expected from \obj's $g-r = 0.64 \pm 0.06$ \citep{Ofek:2012ApJ}.
\obj's $g-r$ places it squarely within the $g-r \sim 0.6-0.7$ colours of other smaller ($H = 4-7$) extreme TNOs, separate from the ultra-red, much larger dwarf planet Sedna ($g-r=0.85, H=1.5$).
Colours similarly just redward of solar are seen for comparably sized $H \sim 6-7$ objects in other dynamically excited TNO populations  
\citep{Sheppard:2010wy,Fraser:2012cs,Peixinho:2015bw,Wong:2017arXiv}.
The colour of \obj\ implies it has a low albedo of $p = 0.05 \pm 0.03$ \citep{Fraser:2014vt,Lacerda:2014}. 

\begin{figure}
\plotone{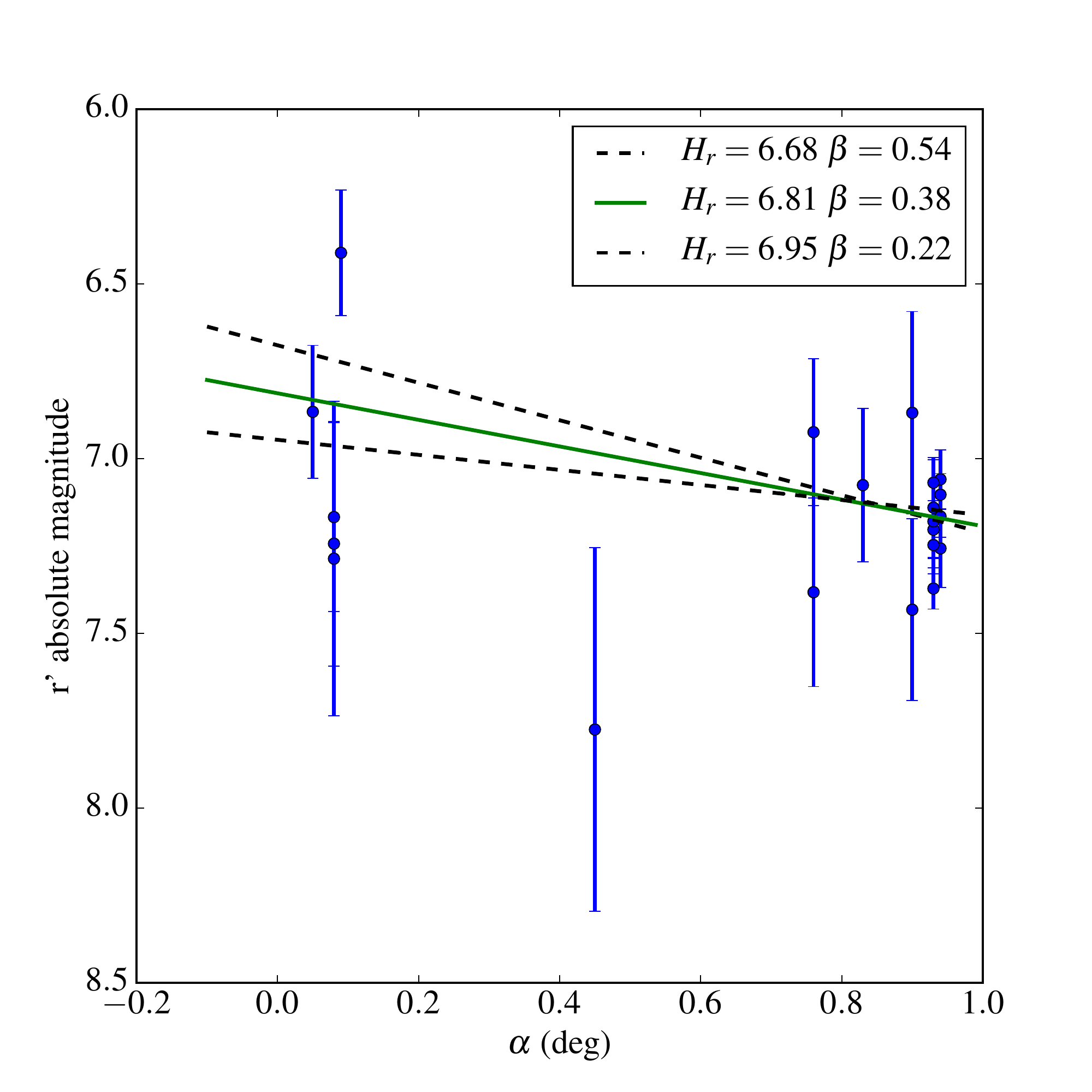}
\caption{The TRIPPy photometry of observations of \obj\ in 2013--2016 from CFHT and Gemini, relative to phase angle (Table~\ref{tab:photometry}). Solid line is the best-fit absolute magnitude $H_r$; dashed lines indicate $1\sigma$ uncertainty in $H_r$ fit.}
\label{fig:phasefit}
\end{figure}

The observations densely sample the ground-based accessible solar phase angle range $0 < \alpha < 1\degr$ (Fig.~\ref{fig:phasefit}). 
This allows for very precise calibration of \obj's intrinsic magnitude, to a level normally achieved for asteroids rather than for TNOs.
We remeasured \obj's photometry with TRIPPy in the CFHT images using the OSSOS-determined centroids and PSF reference stars, with appropriate apertures.
The fluxes measured in Gemini $r$-band were converted into the CFHT $r$-band using the measured colour \citep{Gwyn:2008PASP..120..212G}. The phase and variability behaviour in the combined CFHT-Gemini photometry was fit in a maximum likelihood sense to the observed magnitudes, adopting a linear phase function and sinusoidal light curve. We adopt the median phase curve slope $\beta$, absolute magnitude $H_r$, and lightcurve amplitude $\delta$ reported by \texttt{emcee} \citep{Foreman-Mackey:2013PASP..125..306F} after marginalizing over lightcurve period (considering periods 4--24 hours) and phase reference time. 
\obj\ has $\beta = 0.38\pm0.16$ mag/\degr, consistent with those of other small TNOs \citep{Rabinowitz:2007AJ....133...26R}.
We see variability on few-hour timescales in the Gemini and Subaru data (Table~\ref{tab:photometry}) entirely consistent with our inferred peak-to-peak amplitude $\delta = 0.08$, though we do not attempt to infer a lightcurve as the temporal coverage is insufficient. 
\obj\ has $H_r = 6.81\pm0.14$ and is thus $\sim 250$ km in diameter.

\floattable
\begin{deluxetable*}{lLLll}
\tabletypesize{\footnotesize}
\tablecaption{Published optical SDSS broad-band colours of trans-Neptunian objects with $q > 40$ and $a > 250$ au \label{tab:colours}}
\tablehead{\colhead{Name} & \colhead{$g-r$} & \colhead{$r-i$} & \colhead{Observations}}
\startdata
2012 VP$_{113}$ & 0.70 \pm 0.05 & 0.32 \pm 0.04	& \citet{Trujillo:2014ih} \\
(Sedna) 2003 VB$_{12}$     & 0.85 \pm 0.03	& 0.45 \pm 0.03 & \citet{Sheppard:2010wy} \\
{\bf 2013 SY$_{99}$}		& 0.64 \pm 0.06	& 0.46 \pm 0.10 & {\bf this work}$\dagger$ \\
(474640) 2004 VN$_{112}$ & 0.69 \pm 0.06	& 0.24 \pm 0.06 & \citet{Sheppard:2010wy}  \\
\enddata
\tablecomments{$\dagger$: the two sets of \obj\ colours were obtained on different nights (Table~\ref{tab:photometry}). The $g-r$ colour uses all measurements, weighted by the SNR of each measure.
Given the extended timespan of the Subaru data, the $r-i$ reported is the SNR-weighted colour of the first two $i$ measures, and the first 6 $r$-band, with an 0.05 magnitude uncertainty in quadrature for potential colour terms.
}
\end{deluxetable*}

\section{Orbital integrations} 

\subsection{Present orbital properties}
\label{sec:shortterm}

We examined the allowable range of \obj's current orbit using the algorithm described in \citet{Gladman2008}.
This utilizes a Monte Carlo search for the range of orbits in 6D parameter space that are consistent with the available astrometry, accounting for the possible importance of systematics in the astrometric solution.  
We use a subset of the astrometry, excluding poor-quality images (flagged in Table~\ref{tab:photometry}). 
Because \obj\ was discovered near its perihelion, the available four-opposition arc (\S~\ref{sec:discovery}) allows us to tightly constrain \obj's 50.0 au perihelion distance.
However, this four-opposition, near-perihelion arc is an extremely small fraction of the TNO's near-twenty-thousand-year orbital period, so its $a/e$ combination is relatively weakly constrained.
Very small changes in \obj's difficult to constrain perihelic speed create large changes in $a$.
The Monte Carlo analysis constrains $a$ to be in the range \mina\ to \maxa\ au, 
where the two extremal values in $a$ result in orbit fit residuals no worse than 1.5 times those of the best-fit orbit.
Our uncertainty range is larger than those estimated from the algorithm of \citet{Bernstein:2000p444} because the search allows for systematic residuals.
Other orbital elements are, however, very precisely constrained (Table~\ref{tab:orbits}).

We integrated the best-fit and the extremal orbits forward in time for $10^7$ years with the \texttt{rmvs3} subroutine of SWIFT \citep{Levison1994}.
On a 10 Myr timescale, \obj's semi-major axis changes by more than 1.5 au, which would formally classify it as ``scattering" per the \citet{Gladman2008} criteria.
However, the orbit is metastable on longer Solar System timescales.

\subsection{Long-term evolution in the known Solar System}
\label{sec:longterm}
\begin{figure*}[ht]
\gridline{\fig{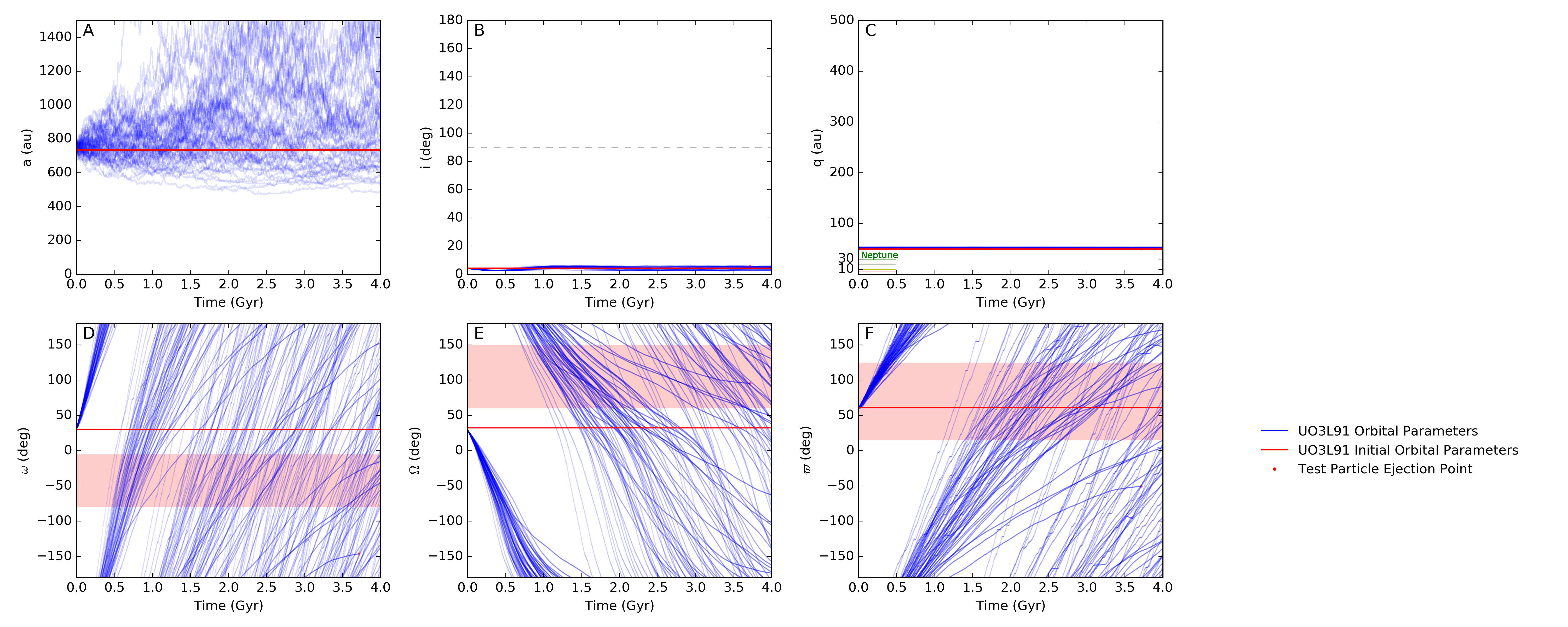}{\textwidth}{(a) Evolution in a Solar System with four giant planets and no Galactic tides. The perihelia of the clones of \obj\ (blue lines) remain constant, while their semi-major axes $a$ diffuse, often out to the inner fringe of the Oort cloud. 
}
}
\gridline{\fig{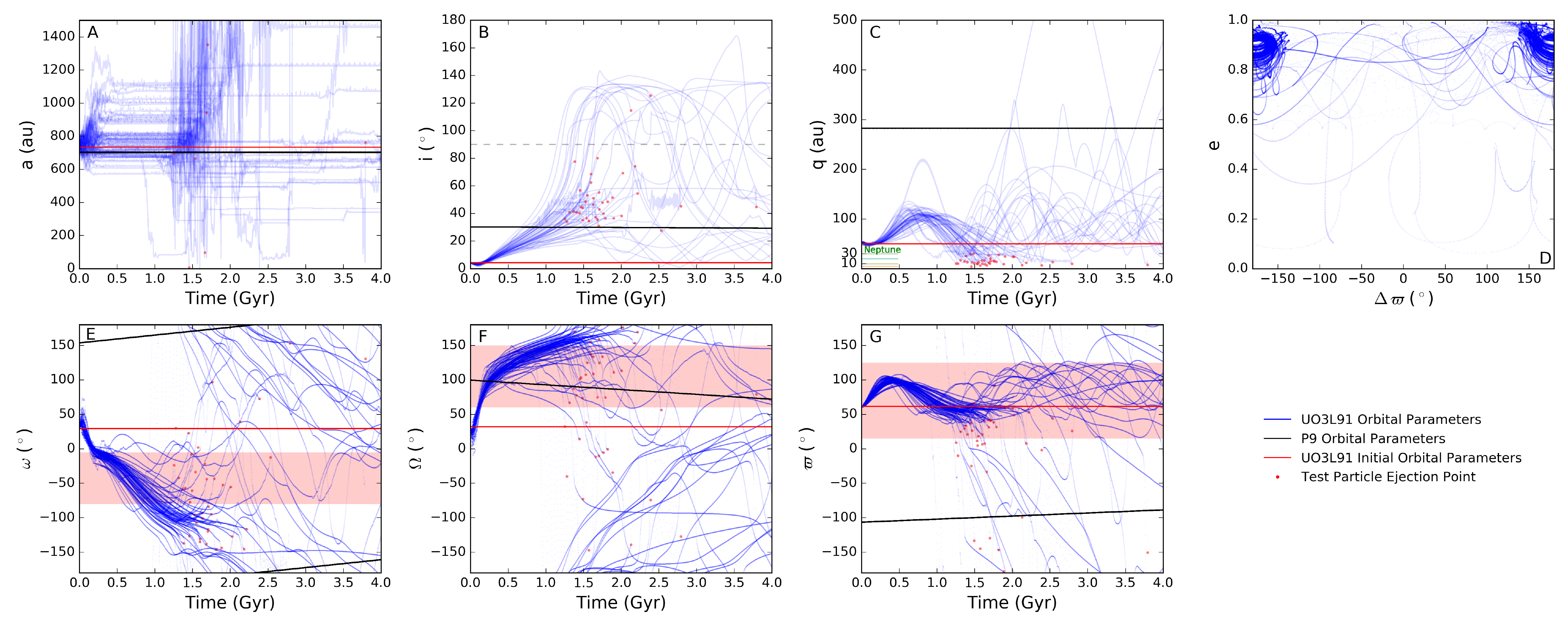}{\textwidth}{(b) Evolution in a Solar System also containing the distant planet of \citet{Batygin:2016ef} ($a=700$ au, $i=30^\circ$, $e=0.6$, $q=280$ au, mass=10 M$_\earth$; black line). A dashed line denotes an inclination of $90\degr$ in panel B. Panel D plots the time evolution of eccentricity versus the difference of longitude of perihelion, $\varpi$, between the additional planet and the test \obj\ clone ($\Delta \varpi = \varpi_{Planet} - \varpi_{clone}$). The clones of \obj\ (blue lines) show substantial instability. For the 33\% that survive ejection by Neptune, perihelion is lifted, while inclination steeply increases toward a perpendicular orbit.}
}
\caption{The orbital evolution over 4 Gyrs for 60 clones of \obj\ randomly sampled within the uncertainty in its orbital parameters in Table~\ref{tab:orbits} (blue lines). Transparent red circles mark the ejection of a test particle in all panels. The best-fit current orbit of \obj\ (red line) and the orbital distances of the four giant planets (lower left corner, Panel C) are indicated in both figures. Pink shading indicates the apparent regions of angular confinement for extreme TNOs as discussed in \citet{Trujillo:2014ih, Batygin:2016ef}; for $\omega$ and $\Omega$, \obj's current parameters fall outside these regions.
}
\label{fig:evolution}
\end{figure*}

We explored the long-term stability of \obj\ in the presence of the four giant planets, both with and without Galactic tides.
Fig.~\ref{fig:evolution} (top) shows the tide-free evolution of 60 clones of \obj\, for 4 Gyr in a 0.5 year time step with the hybrid symplectic/Bulirsch-Stoer algorithm in MERCURY6 \citep{1999MNRAS.304..793C}.
As in \citet{Shankman:2016hi}, three clusters each of 20 clones were generated, at the ($a, q$) extremes and at the nominal orbit, per the orbital uncertainties in Table~\ref{tab:orbits}.
Particles were removed if they reached $a > 10000$ au; this happened to one clone of the 60.

\citet{1987AJ.....94.1330D} considered the root-mean-square (RMS) energy change per perihelion passage for planetary perturbations, as a function of $q$.
For semi-major axes $a \geqslant 100$ au, when the orbit is near parabolic the RMS energy change per encounter is $D(x)$;  
the perihelia stays roughly constant, and the `energy' $x \equiv 1/a$ undergoes a random walk.
The diffusion timescale is $\propto a^{-1/2}$, while $D(x) \ll x$.
Thus the energy changes result in comparatively rapid semi-major axis changes that can be modelled as a random walk or diffusion.
Following \citet{1987AJ.....94.1330D}'s analysis, \obj\ receives weak kicks at perihelion with rms dimensionless energy that are of amplitude $10^{-6}- 3\times10^{-6}$, based on extrapolation of their Fig.~1 out to $q=50$~au for low-$i$ orbits.
For the $a$, $e$, and $i$ of \obj\ we have directly computed that $D(x)\simeq 1.0 \times 10^{-6}$
via numerical scattering experiments, which agrees very well with the extrapolation.
With a perihelion passage every 20 kyr, in 4 Gyr the diffusion should drive an energy walk of fractional amplitude $\delta x = \sqrt{200,000}\ {10^{-6}} x$ producing $\delta a=250$~au, in agreement with Fig.~\ref{fig:evolution}.

The semi-major axis of \obj\ is so large that on Gyr timescales, there is semi-major axis diffusion of 100 au or more.
Thus $a$ can change by a factor of 2 over the age of the Solar System.
Including Galactic tides in our numerical integrations made no appreciable difference to the simulated evolution, as expected per \citet{Kaib:2009Sci}:
while the fraction of clones spending time with $a > 2000$ au is 25\%, the planetary energy kicks still dominate over the effects of Galactic tides, because the $\sim 2$ Gyr tidal torquing time for these orbits is longer. 
The perihelion distance of \obj\ is very stable over time; small oscillations in $q$ are secular, and proportional to semi-major axis, but are never large enough to strongly couple to Neptune.

The diffusion behaviour rules out the possibility of mean-motion resonances (MMR) with Neptune.
Any possible Neptune MMR at $>700$ au would have to be higher than 100th order. 
Such a weak high-order resonance occupies an incredibly thin volume of orbital parameter space: 
the perturbations of \obj's orbit at perihelion would immediately cause a large enough change in $a$ to remove the object from resonance. 

\begin{figure}
\plotone{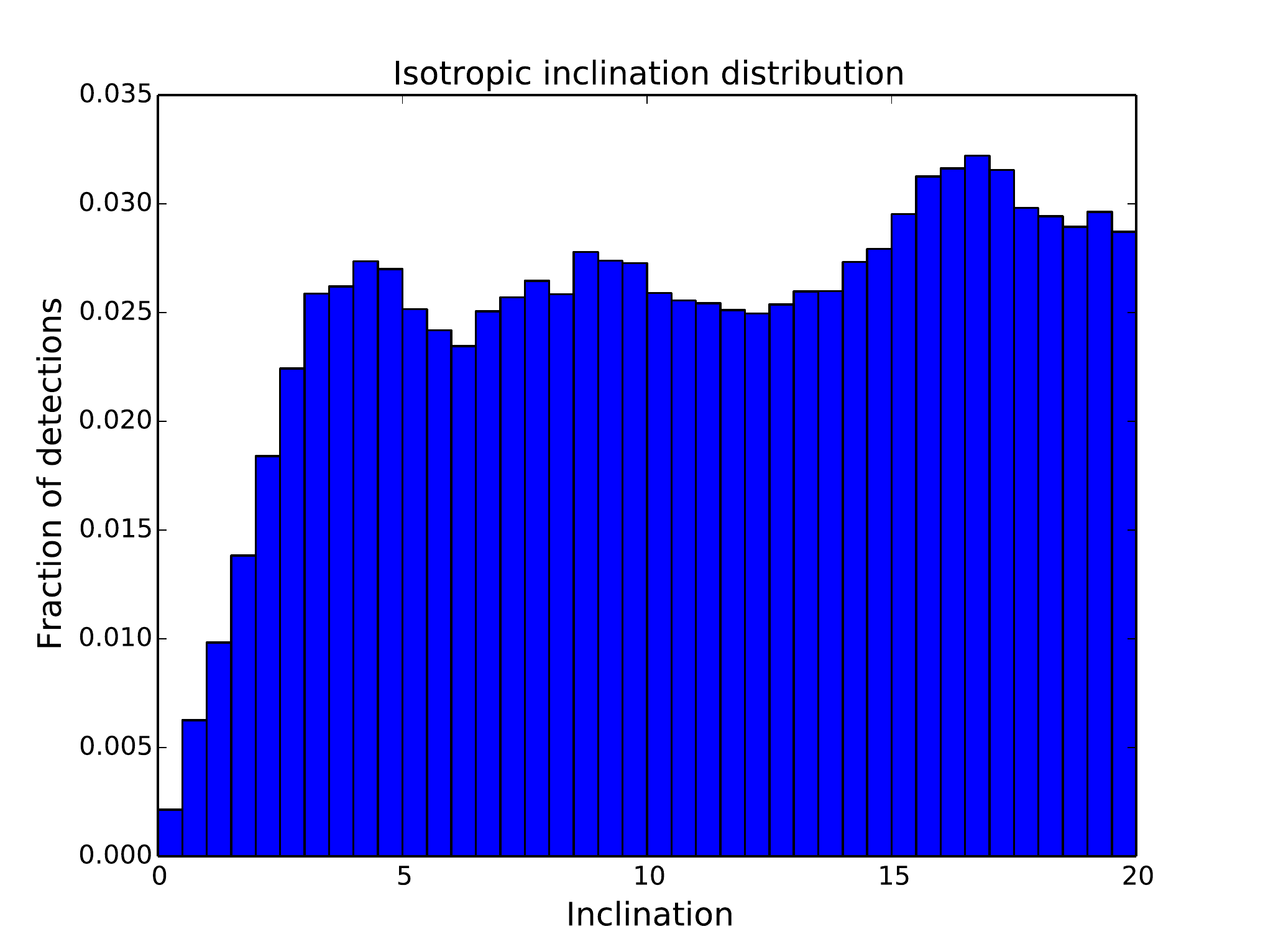}
\caption{The detectability of a set of particles with uniform $100 < a < 1000$ au, $45 < q < 100$ au distributions and a $P(i) \propto \sin(i)$ $0-20\degr$ isotropic inclination distribution, to an ensemble of characterized TNO surveys including the first three sky blocks of OSSOS. \obj\ has $i=4.2\degr$.}
\label{fig:inclination}
\end{figure}

\obj\ has the lowest inclination of the extreme TNOs yet found, with $i = 4.225 \pm 0.001^{\circ}$; the others have inclinations of $12-26\degr$ (Table~\ref{tab:orbits}). 
\obj's inclination remains small: its orbital clones cycled on hundred-Myr time scales between $3-6\degr$.
The survey fields of OSSOS are predominantly in or near the ecliptic, giving the survey more sensitivity to low-$i$ than to high-$i$ orbits.
We used the OSSOS survey simulator \citep{Bannister:2016cp} to test the observability of a population of test particles by an ensemble of well-characterized TNO surveys: OSSOS's 13AE, 13AO and 13BL blocks \citep{Bannister:2016cp}, \citet{Alexandersen:2016ki} and CFEPS/HiLat \citep{Petit:2011p3938,Petit:2016un}.
The particles were given orbits from uniform $100 < a < 1000$ au, $45 < q < 100$ au distributions, with an isotropic inclination distribution from $0-20\degr$ $P(i) \propto \sin(i)$, and random distributions of other angles.  
We drew $H$ magnitudes with $H_r \leq 8$ (recalling $H_{SY99} = 6.8$) from a single slope of $\alpha = 0.9$,
as is consistent with the observed hot and scattering TNO distributions, prior to the transition to an undetermined form past $H_r = 8$ \citep{Fraser:2014vt,Shankman:2016hi}.
The simulated detections, shown in Fig.~\ref{fig:inclination}, are evenly detected at inclinations $3 < i < 20\degr$, dropping at $i < 3\degr$. The $i=4.2\degr$ detection of \obj\ is thus  reasonable for OSSOS. 

\subsection{Diffusion as a way to populate the orbital phase space of extreme TNOs}
\label{sec:diffusion}

The $a$-diffusion exhibited by \obj\ due to planetary perturbations works both ways: as well as causing clones of \obj\ to migrate outward in $a$,
diffusion can be a mechanism to populate \obj's orbital parameter space from more distant regions.
The diffusion pathway was discussed for $31 < q < 65$ au, $a < 500$ au particles by \citet{Gallardo:2012Icar}, and more briefly for $41 < q < 43$ au, $a < 800$ au particles by \citet{Brasser:2014iw}. Neither tested particles with $q \gtrsim 45$ au in larger semi-major axes ranges.

Our initial population of particles were on inner Oort cloud orbits with random inclinations $0 < i < 180\degr$ and $1000<a<2500$ au; $10^4$ particles with $45<q<50$ au,
and a set of smaller batches each of 10 particles with $q$ of 55, 58, 61, 64, 67, 70 au and $1000<a<2000$, all with random distributions of other elements. We evolved these for 4 Gyr under the influence of the giant planets, including the effects of the Galactic tide and passing field stars. 
The orbit end-states are shown in Fig.~\ref{fig:diffusion}. 
After 4 Gyr, most of the particles remain with semi-major axes of thousands of au, with a variety of perihelia with $q > 30$ au. 
13\% have their barycentric distance reach $>200,000$ au and are eliminated from the simulation as ejected. 
The surviving particles with initial perihelia $q \geq 55$ au did not shrink their semi-major axes below their initial $a \geq 1000$ au.
25\% of the particles have semi-major axes that diffuse to $a<2000$ au.
At the 4 Gyr point, 1\% of the particles, all from the initial $45<q<50$ au batch, have orbits with $q > 40$ au and $a < 800$ au. 
The lowest-$a$ $q > 40$ au particle had $a = 306$ au.
This shows diffusion from the inner fringe of the Oort cloud can populate the orbital parameter region of \obj. 

\begin{figure*}
\plottwo{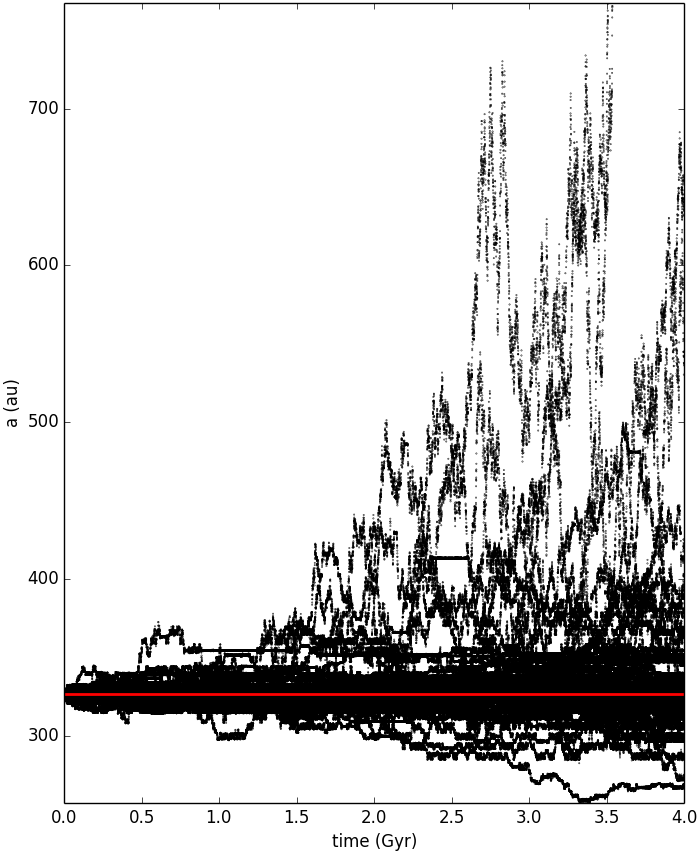}{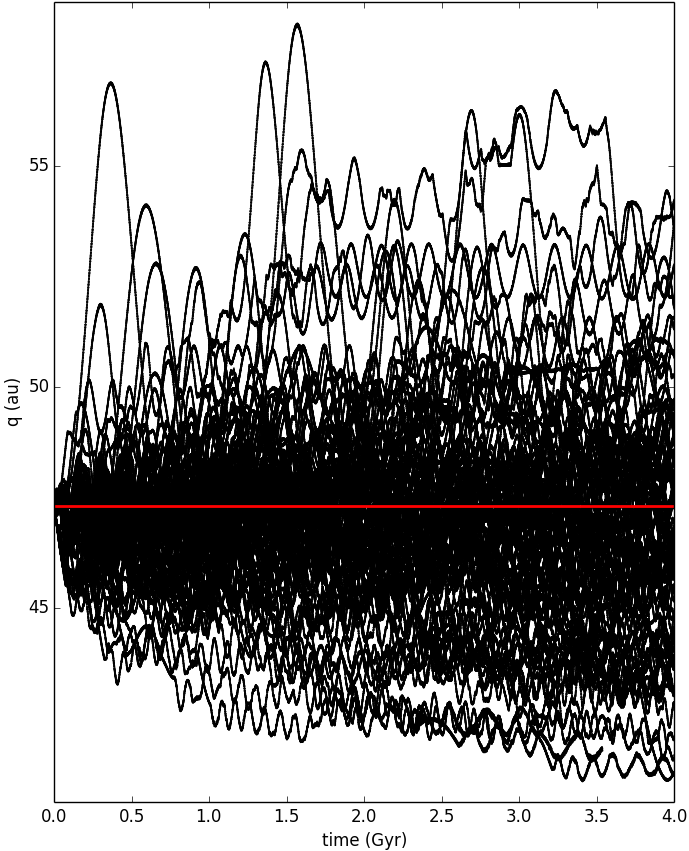}
\caption{The 4 Gyr evolution of semi-major axis and perihelion for 100 clones of (474640) 2004 VN$_{112}$ (black lines) within its phase space, in the presence only of the giant planets. The red line indicates the present best-fit barycentric orbit. This provides an additional example of an extreme TNO exhibiting diffusion.
}
\label{fig:vn112}
\end{figure*}

It is worthwhile to consider if other $45 < q < 50$ au extreme TNOs like 2010 GB$_{174}$ and 2014 SR$_{349}$ (Table~\ref{tab:orbits}) also show diffusive behaviour.
\citet{Sheppard:2016jf} noted semi-major axis mobility for 2013 FT$_{28}$. 
Particles on orbits in the region showed diffusion in the simulations of \citet{Gallardo:2012Icar}; they only sampled part of the phase space occupied by the presently known extreme TNOs. 
We numerically investigate the 4 Gyr orbital evolution of the other extreme TNOs in this $q$ range (including those listed in Table~\ref{tab:orbits}) by generating 100 orbital clones of each object from the orbit-fit covariance matrix; we include only the effects of the Sun and the four giant planets in the simulation. 
Diffusive behaviour in semi-major axis comparable to that of \obj\ is frequent among clones of these extreme TNOs, allowing them to evolve to a similar $a,q$ range. 
An example, (474640) 2004 VN$_{112}$, is shown in Fig.~\ref{fig:vn112}.
These simulations probably underestimate the diffusion and thus should not be used for absolute efficiency estimates of the process at this stage.
We only simulate planetary perturbations; for completeness, further work should also simulate Galactic tides and passing stars. 
We also note that the cloning relies on relatively short arcs from the Minor Planet Center. We encourage further Gaia-calibrated \citep{Lindegren:2016A&A} observations of extreme TNOs by the community. 
All TNOs with clones exhibiting diffusive behaviour are marked in Fig.~\ref{fig:diffusion} (red labels). 
We note that some clones are capable of diffusion, while other clones are stable on Gyr timescales; the process will be more effective for larger-$a$ TNOs, and the phase spaces of some TNOs (e.g. 2014 SR$_{349}$) show less mobility given their current $a, q$.
Sedna and 2012 VP$_{113}$ are not formed by diffusion in the current dynamical environment, consistent with previous assessments \citep{Brown:2004fp,Gallardo:2012Icar,Jilkova:2015gf}.
It is however plausible, for the orbit-fit uncertainty parameter space currently occupied by the orbits of other extreme TNOs, that they could be produced by the same inward diffusion from the inner fringe of the Oort cloud that can populate \obj's parameter region. 

We also find that the clones of some of these lower-$a$ ($a < 400$ au) extreme TNOs show resonance sticking on a variety of timescales from Myr to Gyr. 
We use the definition for resonance set out in \citet{Gladman2008} i.e. libration of a resonant angle for 10 Myr.
Diffusive intervals can lead to periods of sticking to a resonance, particularly for 2014 SR$_{349}$. Resonance sticking was not seen for \obj.
Objects on the lower $a$ and $q$ sides of the ``extreme" region have clones showing stable resonant behaviour with Neptune (Fig.~\ref{fig:diffusion}, blue labels).
For example, 2000 CR$_{105}$ ($a=224.8$ au, $q=44.3$ au) shows insecure occupation in the 20:1 resonance, as long suspected \citep{Gallardo:2006ds}, and some clones of 2013 GP$_{136}$ ($a = 150.2$ au, $q=41.0$ au) librate in a variety of high-order resonances with Neptune.
As seen among the phase space occupied by known extreme TNOs in Fig.~\ref{fig:diffusion}, the processes leading to high-order resonance occupation, and of diffusion, show a gradual overlap between their respective regions of dominance.

\begin{figure*}
\plotone{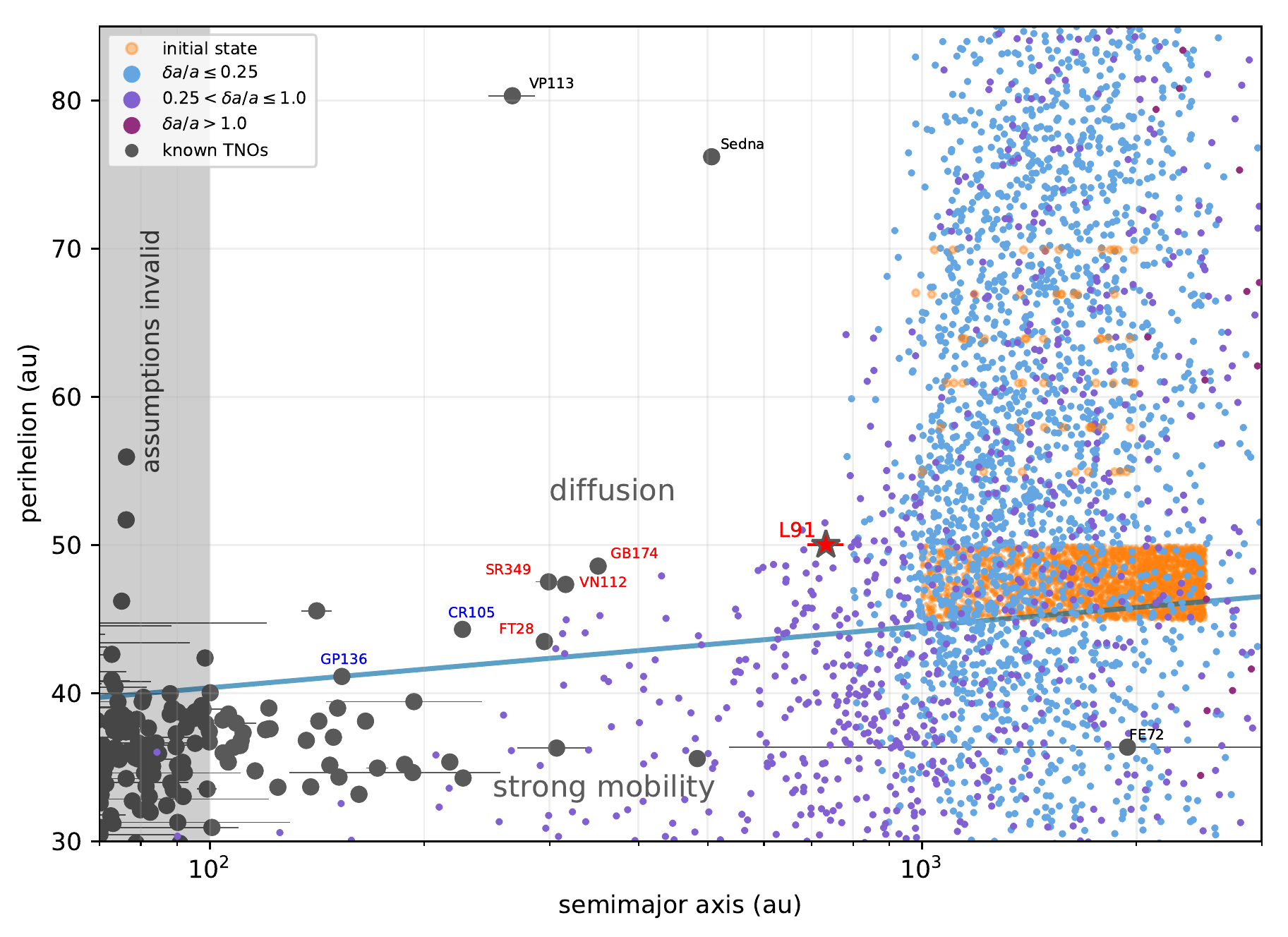}
\caption{The orbits of $\sim10^4$ particles after 4 Gyr (blue/purple dots), begun in confined regions of inner fringe of the Oort cloud orbits (orange dots). Varying degrees of change in $\delta a/a$ after 4 Gyr  are indicated by point colour (blue to purple). Diffusion, driven by weak Neptune interactions, produces the evolution in $a$ that reaches the parameter space where we find \obj\ (red star) and other known extreme TNOs with diffusing orbital clones (red labels). The blue line shows the extrapolated diffusion condition of $\delta a/a = 1$ in 4 Gyr for $i = 0-30\degr$ \citep{1987AJ.....94.1330D}, a rough boundary between diffusive and strongly mobile, actively scattering orbits. At $a < 100$ au the assumptions for diffusive behaviour become invalid (shaded region).
Some known extreme TNOs have clones in high-order mean-motion resonances with Neptune (blue labels).
}
\label{fig:diffusion}
\end{figure*}

\subsection{Behaviour of \obj\ in a Solar System with a distant giant planet}
\label{sec:planet9}

It has recently been argued that the apparently non-random orbital orientations of the few known extreme TNOs (defined using varying criteria) is significant. \citet{Trujillo:2014ih} noted a clustering in the orbits' arguments of pericenter, $\omega$, while \citet{Batygin:2016ef} noted a clustering in the longitude of ascending node, $\Omega$, and in the longitude of pericenter, $\varpi$ (the sum of $\omega$ and $\Omega$), indicating physical alignment of the orbits.  
The orientation of \obj's orbit in relation to the suggested regions of angular confinement is shown in Fig.~\ref{fig:evolution}: \obj\ is external to the previously noted clustering in $\omega$ and $\Omega$, but within that in $\varpi$.
The secular gravitational effects of the giant planets should cause the orbits to precess at different rates, randomizing the orbital angles on $\sim100$ Myr timescales. 

It remains plausible that the complex observational biases in the extreme TNO discovery surveys have strongly influenced the sample \citep{Sheppard:2016jf}.
Extreme TNOs are affected by the usual biases against the discovery of large-$a$ TNOs: their steep luminosity distribution, the $r^{-4}$ drop in reflected solar flux, and their highly eccentric orbits, placing only a fraction of the orbits near perihelion within the observable depth of even large-aperture telescopes. Layered on this is \textit{ephemeris bias}: \citet{Jones:2010bb} noted that high-eccentricity orbits (c.f. \obj's $e = 0.93$) would be most affected by decisions on tracking observations. At discovery (\S~\ref{sec:discovery}), \obj\ was too faint to be tracked as part of our magnitude-limited sample, and its orbit implied it was a commonplace scattering TNO. Only its unusual 60 au detection distance raised the possibility it could potentially be on an interesting high-perihelion orbit.
If \obj\ had instead been at its 50 au perihelion at detection, we would not have worked on tracking it. This illustrates how minor decisions could have biased the known sample. 
Additionally, surveys' sensitivity to the orientation of orbits varies across the sky. \obj\ was detected in an OSSOS block $b = 0-3\degr$ below the ecliptic  (\S~\ref{sec:discovery}). The proximity to the ecliptic induces an observation bias toward detection of objects near perihelion on orbits with argument of perihelion near $0\degr$ or $180\degr$ \citep{Trujillo:2014ih,Sheppard:2016jf}. Unsurprisingly, \obj's orbit has $\omega=32\degr$. 
Lastly, \obj's discovery was at opposition near 1\fh\ R.A., i.e. in September. This time of year (Sept-Dec) experiences typically better seeing (corresponding to deeper limiting magnitudes) in both northern fall and southern spring, continuing a historical seasonal detection bias that can enhance detection at perihelion for surveys, producing clustered longitude of perihelion $\varpi$=0--90\degr.
Thus the simple detection of \obj\ does not yet quantitatively assess whether the apparent extreme TNO clustering arises from a clustered intrinsic population or from observing bias and small-number statistics, and we strongly caution against using it to confirm clustering in the extreme TNOs.
A forthcoming analysis of the full OSSOS survey will quantitatively examine the survey's observing biases for distant TNOs.

If the clustering is a real feature of the population, 
it needs to be forced by a dynamical effect.
Proposed mechanisms are shepherding from a distant massive planet \citep{Trujillo:2014ih, Batygin:2016ef,Brown:2016b}
or self-gravitation in the initial distant planetesimal disk \citep{Madigan:2016iz}. 
\obj\ is problematic for both models: neither predict particles on similar orbits with orbital inclinations as low as that of \obj.

We test the long-term dynamical behaviour of \obj\ as in \S~\ref{sec:longterm}, with the addition of the inclined $i = 30\degr$ $a = 700$ au 10 M$_\earth$ planet of \citet{Batygin:2016ef} (Fig.~\ref{fig:evolution}, lower).
This simulation does not incorporate tides, as the perturbations due to the planet are orders of magnitude greater \citep{Shankman:16B}.
The effect of the additional planet is dramatic: on hundred-Myr timescales, \obj's perihelion distance undergoes significant cyclic variation. For most clones, $q$ drops low enough for the object to experience close encounters with the known giant planets, and is ejected from the Solar System. 
In contrast to the complete stability seen in the known planetary configuration, \added{40 of the 60} \obj\ clones are ejected within a Gyr in a Solar System with an additional planet.
For the surviving 33\% of the clones of \obj, their perihelion distances are drawn outward while their inclinations steeply increase, until the particles are orbiting perpendicular to the plane of the Solar System.
The inclination and perihelia of the surviving particles cycle widely, on near-Gyr timescales; these clones also exhibit $\varpi$ shepherding as in \citet{Batygin:2016ef}.
The $i$ and $q$ evolution of the clones would sharply reduce the detectability of \obj, \added{as the clones spend almost all their time at high inclinations and larger perihelia}. 
Similar behaviour is seen by \citet{Shankman:16B} for other extreme TNOs under the \citet{Batygin:2016ef} planet parameters.
The removal from the detectable volume and high ejection fraction of \obj\ clones thus requires an enormous population of cycling TNOs to permit the detection of \obj, potentially an $M_\earth$ or more. High mass estimates are also found in the simulations of \citet{Lawler:2016fg}.

With appropriate external planet parameters, \obj\ could be phase-protected in a mean-motion resonance \citep{Batygin:2016ef, Malhotra:2016ab}, though this is not strictly required \citep{Beust:2016ix}.
However, the present orbit of \obj\ is too uncertain to discuss potential resonances with an external planet as in \citet{Malhotra:2016ab,DlFM2:2016MNRAS.460L..64D,Millholland:2017er}. 
Several years of further observation will be needed to constrain the semi-major axis to within the width of a potential resonance. 
There are substantial compounding uncertainties: the technically infinite number of potential resonances, the widths of resonances, the $a$ uncertainties in the extreme TNO sample, and the flexible range for the potential planet's semi-major axis.
Any potential proximity even to low-order resonances is thus currently not significant.

\section{Discussion}

We consider how many objects must exist in the orbital parameter space of \obj\ for us to have detected it.
\obj\ was substantially fainter than the characterization limit at discovery (\S~\ref{sec:discovery}). As such, its 9\% detection efficiency is only roughly estimated, in contrast to the well-constrained efficiencies for the vast majority of OSSOS discoveries, and a derived population estimate must have a factor of ten uncertainty.
Proceeding then with caution, the observed existence of \textit{any} extreme TNOs still implies the presence of a vast unseen reservoir \citep{Gladman:2002Icar, Schwamb:2010p932,Kaib:2009Sci,Chen:2013gs}.

No model exists for the formation of this population, and the sample size is too small to have a reasonably constrained parametric model. 
We instead adopt a population deliberately limited in scope, which will be indicative of an underlying population, but cannot be robustly tested for rejectability.

We simulated the observability of an \obj-like set of particles to an ensemble of surveys\footnote{We also made an independent test with the Next Generation Virgo Survey \citep{Chen:2013gs}, which did not provide any stricter constraints.}, as in \S~\ref{sec:longterm}.
The test particle population was a uniformly filled $a-e-i$ box around the uncertainties in the orbit of \obj, assigned random orbit angles and an $H$ magnitude. We used a single slope luminosity function of $\alpha = 0.9$ down to $H_r = 8$, a choice justified in \S~\ref{sec:longterm}.
\obj's 0.1 mag variability (\S~\ref{sec:discovery}) can only shift the population by a factor of 1.5.
We selected random particles from the model until one was detectable.
For us to detect \obj, there need to exist $5 \times 10^5$ $H_r < 8$ objects of $q \sim 50$ and large $a$, to order of magnitude precision.
This is consistent with the estimate of \citet{Gladman:2002Icar}.
Given \obj's albedo (\S~\ref{sec:discovery}), this population with $H_r < 8$ would have a mass of $0.04 \pm 0.01 M_\earth$.
\citet{Trujillo:2014ih}\footnote{The more extensive continued survey in \citet{Sheppard:2016jf} does not provide a population estimate for a full range of angles for the orbit ranges discussed here, only an asymmetric clustered-population estimate.}, inferred $0.03 M_\earth$ for a symmetric and larger orbital parameter range $q > 50, 200 < a < 800$ au population with $\alpha=0.9$. 

There are three possibilities for the extreme population: 
\begin{itemize}
\item it is in steady state, after primordial emplacement of the inner fringe of the Oort cloud; 
\item it was an early emplacement that is decaying over the age of the Solar System, like the scattered disk; 
\item such objects act like Centaurs, continuously scattered through and rapidly removed from the space between Neptune and a potential outer planet.
\end{itemize}
We do not assess the last case here.
In the steady-state and decaying cases, our simulations in \S~\ref{sec:diffusion} imply $\sim 1$\% of the inner fringe of the Oort cloud have diffused inward and are present in $q>40, 300 < a < 800$ au parameter space. The nature of diffusion means the population are actively, on long time scales, cycling back and forward.
We encourage further investigations of the efficiency of the diffusion process, as it could directly probe the mass resident in the inner fringe of the Oort cloud.
For initial emplacement of the population, \citet{Dones:2004} found that the giant planets scatter material from the initial protoplanetesimal disk into the whole Oort cloud with an efficiency of $\sim 3\%$. \citet{Kaib:2011ky} find $\sim 10\%$ of the Oort cloud mass in the 1000-2000 au region, so the scattering efficiency to there should be a few tenths of a percent. Diffusion as a solo process may thus imply an exceptionally massive inner fringe; however, it is unlikely that any one dynamical process solely modifies the population across all Solar System history.
 
We consider the extreme-TNO origin scenarios to see if any formation models for \obj\ can be ruled out.
A formation scenario must account for both \obj's detached perihelion and its large semi-major axis.

The \textit{inclination instability} of self-gravitating planetesimals \citep{Madigan:2016iz} cannot yet be excluded; it requires a mass of several M$_\earth$ of objects orbiting with $100 < a < 10,000$ au, which is possible given the population \obj\ implies in only $a < 2000$ au.

A presently orbiting \textit{distant planet} has the advantage of raising perihelia, though we disfavour it due to the \replaced{instability and loss of observability of \obj}{loss of $\sim 70\%$ of \obj\ particles and removal of the retained $\sim 30\%$ of particles from the observationally detectable space,} seen in \S~\ref{sec:planet9}.
Scattering from \textit{lost embryos/planets} remains plausible. Particles covering the orbital parameter space of all extreme TNOs are produced in the rogue planet simulations of \citet{2006ApJ...643L.135G}, including the decreasing maximum inclination trend with semi-major axis of $i_{imax}=70^{\circ} \times ($100 au$/a)$ present in the observed sample.

Other plausible perihelia-raising mechanisms are \textit{perturbation in the Sun's birth cluster}, specifically the cluster potential \citep{Jilkova:2015gf},
and following the departure of the Solar System from the birth cluster, \textit{stellar encounters}, which can emplace a few tenths of $M_\earth$ \citep{2004AJ....128.2564M}.
\citet{Kaib:2011ky} found in simulations that only include the standard Galactic tide and passing solar neighbourhood stars, minor planets have their orbital perihelia torqued out of the planetary region with $a\sim 1000$ au. Stellar perturbations are a random process, and the most powerful individual stellar encounters the Sun will experience during 4 Gyrs provide much more powerful perihelia torquing than the Galactic tide (on the timescale of the encounter or on the Myr timescale of an $a$-diffusing orbit, which is short compared to 4 Gyrs). 

We outline a \textit{scattering to diffusion} scenario for the formation of \obj\ that is entirely compatible with the known Solar System.
An object scatters outward in the initial emplacement of the scattering disk, pushing the semi-major axis of its orbit into the inner fringe of the Oort cloud (e.g. 2014 FE$_{72}$) \citep{Gladman:2005Sci...307...71G,2006Icar..184..619L}.
At a semi-major axis of a thousand or more au, Galactic tides couple and torque out the orbit's perihelion.
Once an object is orbiting with $q=50$ au and $a\sim 1000-2000$ au, it diffuses to a lower-$a$ orbit via planetary energy kicks.
A reservoir population of objects must then exist that cycles under diffusion with $q\sim 40-50$ au and $a \sim 1000-2500$ au. 
This dynamical pathway is possible in the simplest Oort Cloud model.
The shared neutral colour of \obj\ and similar-sized objects both in the scattering disk and diffusing extreme TNOs also potentially support a common ancestral origin point in the giant planet region.

We emphasize that this scenario can form the larger-$a$ $q\sim 40-50$ au extreme TNOs, but does not form the $q \sim 80$ au TNOs like Sedna: there will be a maximum perihelion above which the kicks from Neptune that permit diffusion to lower semi-major axis become too weak.
It is then worthwhile to consider the perihelia distribution produced by the extreme TNO formation scenarios.
Models of stellar perturbation in a cluster environment form Sedna and other $q \sim 80$ au objects as a generic outcome independent of cluster size, with an inner edge to the $q$ distribution \citep{Brasser:2012gi}, which had to be $q \geq 75$ au \citep{Schwamb:2010p932,Sheppard:2016jf}.
Extreme TNOs with perihelia greater than an initial $q_0$ are formed by the capture of extra-solar planetesimals during the Sun's cluster birth \citep{Levison:2010Sci,Jilkova:2015gf}.
In contrast, a distant planet would smoothly lift the perihelia of the extreme TNOs \citep{Gomes:2006en,Lawler:2016fg, Shankman:16B}. 
In our scenario, the initial outward draw of perihelia by tides coupling to large-$a$ scattering objects would also smoothly distribute perihelia.

Fig.~\ref{fig:diffusion} shows the current absence of extreme TNOs with orbital perihelia in the range 50-75 au \citep{Trujillo:2014ih,Sheppard:2016jf}. 
This ``perihelion gap" remains, given \obj\ has $q = 50.0$ au. 
A gap in the population excludes the existence of a presently orbiting planet, which would otherwise be cycling minor planets through this region \citep{Shankman:16B}. It also rules out other mechanisms for lifting perihelia.
OSSOS has found small TNOs at barycentric distances of 50--75 au, and the OSSOS survey simulator indicates TNOs with perihelia 50--75 au would have been detectable. Testing if their absence is merely a discovery bias is harder; that requires the development of a population model that completely links both this region and another better-understood population like the scattering disk. 
However, our scenario for forming \obj's orbit does show that for an inner Oort cloud object with $q$ lifted to $\gtrsim 55$ au, diffusion will be too weak to retract the semi-major axis (Fig.~\ref{fig:diffusion}). Thus future discoveries with $q \sim 60$ au should have $a \gtrsim 1000$ au.

\section{Conclusion}

Our discovery of an $a = \besta \pm \sigmaa$ au, $q = 50.0$ trans-Neptunian object shows the extreme detached TNO population occupies orbital phase space with much larger semi-major axes than previously seen.
We model its orbital evolution over 4 Gyr and find that weak planetary kicks from Neptune at its perihelia still link \obj, and the phase space of other extreme TNOs, to planetary perturbation.
Via known processes in the current Solar System, TNOs on orbits with $q \sim 45-50$ au and $a \sim 700$ au can be pulled down from semi-major axes beyond 1000 au in the inner fringe of the Oort cloud, and move back to similarly large semi-major axes.

We propose a scenario where minor planets scattered outward to the inner fringe of the Oort cloud as part of the emplacement of the scattered disk, Galactic tides lifted the perihelion distances of objects in the inner fringe of the Oort cloud, and diffusion walked their semi-major axes back, forming the population of \obj.
This process should form an ongoing cycling of small worlds between the orbital parameter space of \obj\ and the $a = 1000-2000$ region.
The colour of \obj\ is entirely consistent with a shared primordial population.
\obj\ is thus the closest minor planet to an inner Oort cloud object yet seen.
Its existence further supports a substantial $a \sim 1000-2000$ au reservoir of small bodies.

\acknowledgments

The authors recognize and acknowledge the sacred nature of Maunakea, and appreciate the opportunity to observe from the mountain. 
This project could not have been a success without the dedicated staff of the Canada--France--Hawaii Telescope (CFHT). 
CFHT is operated by the National Research Council of Canada, the Institute National des Sciences de l'Univers of the Centre National de la Recherche Scientifique of France, and the University of Hawaii, 
with OSSOS receiving additional access due to contributions from the Institute of Astronomy and Astrophysics, Academia Sinica, Taiwan.
This work is based on observations obtained with MegaPrime/MegaCam, a joint project of CFHT and CEA/DAPNIA; on data produced and hosted at the Canadian Astronomy Data Centre and on the CANFAR VOSpace; on observations obtained at the Gemini Observatory, which is operated by the Association of Universities for Research in Astronomy, Inc., under a cooperative agreement with the NSF on behalf of the Gemini partnership: the National Science Foundation (United States), the National Research Council (Canada), CONICYT (Chile), Ministerio de Ciencia, Tecnolog\'{i}a e Innovaci\'{o}n Productiva (Argentina), and Minist\'{e}rio da Ci\^{e}ncia, Tecnologia e Inova\c{c}\~{a}o (Brazil); and on data collected at Subaru Telescope, which is operated by the National Astronomical Observatory of Japan.
M.T.B. acknowledges support from UK STFC grant ST/L000709/1, the National Research Council of Canada, and the National Science and Engineering Research Council of Canada.
M.J. acknowledges the support of the Slovak Grant Agency for Science (grant VEGA No. 2/0031/14).

\facility{CFHT (MegaPrime), Gemini, Subaru}.
\software{Python, astropy, trippy, matplotlib, scipy, numpy, emcee, Ureka, SWIFT, MERCURY6}

\appendix
\section{Photometry of \obj}
\label{sec:colours_lc}

\startlongtable
\begin{deluxetable*}{ccccLc}
\tablecaption{Observations of \obj\ with CFHT, Gemini and Subaru \label{tab:photometry}}
\tablehead{
\colhead{MPC} & \colhead{Time} & \colhead{Filter} & \colhead{Exposure} & \colhead{Magnitude} & \colhead{Solar phase} \vspace{-0.25cm} \\
\colhead{flag} & \colhead{(Midpoint, UT)} & \colhead{} & \colhead{Time (s)} & \colhead{in passband} & \colhead{$\alpha$ ($\degr$)} } 
\startdata
\cutinhead{CFHT MegaCam: (photometry: OSSOS, $\dagger$-flagged: TRIPPy)}
  & 2013 09 05.42994 & R.3901 & 287 & 25.28 \pm 0.52\dagger & 0.45	 \\ 
V & 2013 09 05.46880 & R.3901 & 287 & 24.74 \pm 0.33 & 0.45	 \\ %
$*$ & 2013 09 29.36094 & R.3901 & 287 & 24.76 \pm 0.27 \dagger & 0.08 \\ 
  & 2013 09 29.39977 & R.3901 & 287 & 24.98 \pm 0.45 \dagger & 0.08	 \\ 
d & 2013 09 29.44359 & R.3901 & 287 & 24.09 \pm 0.18 & 0.08	 \\ %
V & 2013 09 29.44819 & R.3901 & 287 & 24.71 \pm 0.35 \dagger & 0.08	 \\ 
d & 2013 09 29.45297 & R.3901 & 287 & 24.22 \pm 0.23 & 0.08	 \\ %
V & 2013 09 29.45785 & R.3901 & 287 & 24.00 \pm 0.18 & 0.08	 \\ %
d & 2013 09 29.46414 & R.3901 & 287 & 24.02 \pm 0.22 & 0.08	 \\ %
d & 2013 09 29.46913 & R.3901 & 287 & 24.26 \pm 0.29 & 0.08	 \\ 
H & 2013 09 29.50184 & R.3901 & 287 & 24.98 \pm 0.39 & 0.08	 \\ 
V & 2013 10 05.35958 & R.3901 & 287 & 25.41 \pm 0.50 & 0.03	 \\ 
V & 2013 10 06.36797 & R.3901 & 287 & 24.23 \pm 0.19 \dagger & 0.05	 \\ 
V & 2013 10 06.43244 & R.3901 & 287 & 25.25 \pm 0.38 & 0.05	 \\ 
I & 2013 10 07.35479 & R.3901 & 287 & 23.68 \pm 0.15 & 0.06	 \\ 
d & 2013 10 09.49249 & R.3901 & 287 & 23.94 \pm 0.18 \dagger & 0.09	 \\ 
d & 2013 11 27.29243 & R.3901 & 287 & 24.84 \pm 0.27 \dagger & 0.76	 \\ 
  & 2013 11 27.33867 & R.3901 & 287 & 24.43 \pm 0.21 \dagger & 0.76	 \\ 
V & 2013 12 05.24094 & R.3901 & 300 & 24.55 \pm 0.22 \dagger & 0.83	 \\ 
V & 2014 12 17.28612 & R.3901 & 500 & 24.28 \pm 0.29 \dagger & 0.90	 \\ 
  & 2015 01 17.25897 & R.3901 & 387 & 24.94 \pm 0.26 \dagger & 0.90	 \\ 
  & 2015 10 09.40506 & GRI.MP9605 & 300 & 24.53 \pm 0.13 \dagger & 0.90	 \\ 
  & 2015 10 09.44841 & GRI.MP9605 & 300 & 24.97 \pm 0.15 \dagger & 0.90	 \\ 
  & 2015 10 09.49091 & GRI.MP9605 & 300 & 24.59 \pm 0.09 \dagger & 0.90	 \\ 
H & 2015 11 17.26167 & GRI.MP9605 & 450 & - 	 \\
  & 2015 12 07.39899 & GRI.MP9605 & 400 & 24.79 \pm 0.15 \dagger & 0.84	 \\ 
  & 2015 12 31.22796 & GRI.MP9605 & 400 & 24.63 \pm 0.10 \dagger & 0.90	 \\ 
  & 2016 07 07.58727 & GRI.MP9605 & 300 & 24.81 \pm 0.08 & 0.98	 \\ %
  & 2016 07 08.56666 & GRI.MP9605 & 300 & 24.67 \pm 0.12 & 0.98	 \\ %
  & 2016 07 08.57071 & GRI.MP9605 & 300 & 24.74 \pm 0.13 & 0.98	 \\ %
d & 2016 07 08.57475 & GRI.MP9605 & 300 & 24.88 \pm 0.14 & 0.98	 \\ %
  & 2016 11 04.31217 & GRI.MP9605 & 450 & 24.66 \pm 0.09 & 0.46  \\ %
I & 2016 11 04.37124 & GRI.MP9605 & 450 & 24.46 \pm 0.06 & 0.46	 \\ %
\cutinhead{Gemini GMOS-N (photometry: TRIPPy)}  
  & 2016-01-07 05:04:57	& r.G0303 &	600  & 24.7069 \pm 0.0572 & 0.94 \\
  & 2016-01-07 05:15:34	& r.G0303 &	600  & 24.6452 \pm 0.0581 & 0.94 \\ 
  & 2016-01-08 05:11:22	& r.G0303 &	600  & 24.6145 \pm 0.0831 & 0.94 \\
  & 2016-01-08 05:21:58	& r.G0303 &	600  & 24.8176 \pm 0.1110 & 0.94 \\ 
  & 2016-01-11 04:45:51	& r.G0303 &	600  & 24.6935 \pm 0.1419 & 0.93 \\
  & 2016-01-11 04:56:28	& r.G0303 &	600  & 24.6477 \pm 0.0635 & 0.93 \\
  & 2016-01-11 05:07:06	& r.G0303 &	600  & 24.8037 \pm 0.0566 & 0.93 \\
  & 2016-01-11 05:17:50	& g.G0301 &	600  & 25.2567 \pm 0.0769 & 0.93 \\
  & 2016-01-11 05:28:27	& g.G0301 &	600  & 25.3249 \pm 0.0850 & 0.93 \\
  & 2016-01-11 05:39:04	& g.G0301 &	600  & 25.4567 \pm 0.0998 & 0.93 \\
  & 2016-01-11 05:49:42	& g.G0301 &	600  & 25.4985 \pm 0.1061 & 0.93 \\
  & 2016-01-11 06:00:19	& g.G0301 &	600  & 25.5788 \pm 0.1153 & 0.93 \\
  & 2016-01-11 06:10:57	& g.G0301 &	600  & 25.5309 \pm 0.1145 & 0.93 \\
  & 2016-01-11 06:21:32	& g.G0301 &	600  & 25.2420 \pm 0.0913 & 0.93 \\
  & 2016-01-11 06:32:09	& g.G0301 &	600  & 25.3515 \pm 0.0957 & 0.93 \\
  & 2016-01-11 06:42:47	& g.G0301 &	600  & 25.6312 \pm 0.1164 & 0.93 \\
  & 2016-01-11 06:53:22	& g.G0301 &	600  & 25.4617 \pm 0.1069 & 0.93 \\
  & 2016-01-11 07:04:05	& r.G0303 &	600  & 24.7828 \pm 0.0810 & 0.93 \\
  & 2016-01-11 07:14:42	& r.G0303 &	600  & 24.8486 \pm 0.0814 & 0.93 \\
  & 2016-01-11 07:25:20	& r.G0303 &	600  & 24.7414 \pm 0.1049 & 0.93 \\
\cutinhead{Subaru Hyper Suprime Cam (photometry: TRIPPy)}
  & 2016-01-10 05:13:24 & HSC-I & 300.0  & 24.6001 \pm 0.0939 & 0.94 \\
  & 2016-01-10 05:18:56 & HSC-I & 300.0  & 24.5272 \pm 0.1104 & 0.94 \\
  & 2016-01-10 05:52:21 & HSC-R & 300.0  & 25.0422 \pm 0.0977 & 0.94 \\
  & 2016-01-10 05:57:54 & HSC-R & 300.0  & 25.3657 \pm 0.1394 & 0.93 \\
  & 2016-01-10 06:03:30 & HSC-R & 300.0  & 25.0390 \pm 0.1285 & 0.93 \\
  & 2016-01-10 06:09:02 & HSC-R & 300.0  & 25.0496 \pm 0.1157 & 0.93 \\
  & 2016-01-10 06:14:34 & HSC-R & 300.0  & 24.9846 \pm 0.0952 & 0.93 \\
  & 2016-01-10 06:20:12 & HSC-R & 300.0  & 24.8008 \pm 0.1086 & 0.93 \\
  & 2016-01-10 07:49:23 & HSC-R & 300.0  & 24.6847 \pm 0.1009 & 0.93 \\
  & 2016-01-10 07:54:58 & HSC-R & 300.0  & 24.8982 \pm 0.1622 & 0.93 \\
  & 2016-01-10 08:00:31 & HSC-R & 300.0  & 24.7119 \pm 0.1097 & 0.93 \\
\enddata
\end{deluxetable*}

\bibliography{highq}
\bibliographystyle{aasjournal}

\end{document}